%% file: main.tex
\documentclass[trackchanges,twocolumn]{aastex7}

\usepackage{xcolor}
\usepackage{xspace}

\definecolor{allysoncolor}{RGB}{0, 128, 255} 
\definecolor{karencolor}{RGB}{180, 0, 255}

\newcommand\tierrasenddate{2025~May~30}
\newcommand\ntierrassources{75}
\newcommand\tierraslsper{11.00}

\newcommand\tierrasmcmcprot{\ensuremath{11.020 \pm 0.015}}
\newcommand\tierrasmcmclambda{\ensuremath{{305.1^\circ}^{+5.4^\circ}_{-5.3^\circ}}}
\newcommand\tierrasmcmcpsi{\ensuremath{{77.4^\circ} ^{+2.3^\circ}_{-2.5^\circ}}}
\newcommand\tierrasmcmcrspot{\ensuremath{{31.2^\circ}^{+2.4^\circ}_{-1.9^\circ}}}
\newcommand\tierrasmcmclatspot{\ensuremath{{80.5^\circ}\pm 1.2^\circ}}
\newcommand\tierrasmcmcistar{\ensuremath{{22.3^\circ}^{+1.8^\circ}_{-1.6^\circ}}}

\newcommand{\gcc}{\ensuremath{\mathrm{g}\,\mathrm{cm}^{-3}}\xspace}
\newcommand{\kms}{\ensuremath{\mathrm{km}\,\mathrm{s}^{-1}}\xspace}
\newcommand{\ms}{\ensuremath{\mathrm{m}\,\mathrm{s}^{-1}}\xspace}

\usepackage{makecell}

\defcitealias{Almenara2022}{A22}
\defcitealias{LibbyRoberts2023}{LR23}

\begin{document}

\title{Spot-Crossing Variations Confirm a Misaligned Orbit for a Planet Transiting an M Dwarf}

\author[0000-0003-2171-5083]{Patrick Tamburo}
\affiliation{Center for Astrophysics $\vert$ Harvard \& Smithsonian, 60 Garden Street, Cambridge, MA 02138, USA}
\email[show]{patrick.tamburo@cfa.harvard.edu}  

\author[0000-0001-7961-3907]{Samuel W. Yee}\altaffiliation{51 Pegasi b Fellow}
\affiliation{Center for Astrophysics $\vert$ Harvard \& Smithsonian, 60 Garden Street, Cambridge, MA 02138, USA}
\email[]{samuel.yee@cfa.harvard.edu}  

\author[0000-0003-1361-985X]{Juliana Garc\'ia-Mej\'ia}\altaffiliation{51 Pegasi b Fellow}
\affiliation{Center for Astrophysics $\vert$ Harvard \& Smithsonian, 60 Garden Street, Cambridge, MA 02138, USA}
\affiliation{Kavli Institute for Astrophysics and Space Research, Massachusetts Institute of Technology, Cambridge, MA 02139, USA}
\email[]{juliana.garcia-mejia@cfa.harvard.edu}  

\author[0000-0002-9003-484X]{David Charbonneau}
\affiliation{Center for Astrophysics $\vert$ Harvard \& Smithsonian, 60 Garden Street, Cambridge, MA 02138, USA}
\email[]{dcharbonneau@cfa.harvard.edu}  

\author[0000-0001-6637-5401]{Allyson~Bieryla}
\affiliation{Center for Astrophysics $\vert$ Harvard \& Smithsonian, 60 Garden Street, Cambridge, MA 02138, USA}
\email[]{abieryla@cfa.harvard.edu} 

\author[0000-0001-6588-9574]{Karen~A.~Collins}
\affiliation{Center for Astrophysics $\vert$ Harvard \& Smithsonian, 60 Garden Street, Cambridge, MA 02138, USA}
\email[]{karen.collins@cfa.harvard.edu}

\author[0000-0002-1836-3120]{Avi Shporer}
\affiliation{Kavli Institute for Astrophysics and Space Research, Massachusetts Institute of Technology, Cambridge, MA 02139, USA}
\affiliation{Department of Physics, Massachusetts Institute of Technology, Cambridge, MA 02139, USA}
\email[]{shporer@mit.edu}

\begin{abstract}
TOI-3884~b is an unusual 6.4~R$_\oplus$ planet orbiting an M4 host, whose transits display large and persistent spot-crossing events. We used the \textit{Tierras} Observatory to monitor both the long-term photometric variability of TOI-3884 and changes in the spot-crossing events across multiple transits of the planet. We show that the star rotates with a period of \tierrasmcmcprot~days. We simultaneously model the rotational modulation of the star and variations in transit shapes that arise due to rotation of the spot, allowing us to determine the true stellar obliquity, $\psi_\star$. The data are best described by a planet on a misaligned orbit around a highly inclined star ($\psi_\star = $~\tierrasmcmcpsi; $i_\star = $~\tierrasmcmcistar) that hosts a large polar starspot ($r_\mathrm{spot} = $~\tierrasmcmcrspot; $\lambda_\mathrm{spot} = $~\tierrasmcmclatspot). Archival photometry from the Zwicky Transient Facility suggests that this polar spot has persisted on TOI-3884 for at least seven years. The TOI-3884 system provides a benchmark for studying the evolution of a polar spot on an M dwarf.  
\end{abstract}



\section{Introduction} \label{sec:intro}

TOI-3884 is an M4 dwarf star ($T_\mathrm{eff} = 3180$~K, $R_\star = 0.30~R_\odot$, $M_\star = 0.30~M_\odot$) that hosts a 6.4~$R_\oplus$, 33~$M_\oplus$ transiting planet on a 4.5-day orbit (\citealt{Almenara2022}; \citealt{LibbyRoberts2023}; hereafter \citetalias{Almenara2022} and \citetalias{LibbyRoberts2023}). Planets this large are rare around such low-mass stars. \citet{Pass2023b} found that fewer than 7.1\% of mid-to-late M dwarfs host planets with $ 31.8~M_\oplus < M_\mathrm{p}\sin{i} < 56.5~M_\oplus,~ 3 < P < 10 $~days (95\% confidence limit), based on an RV survey of a 15~pc volume-complete sample \citep{Winters2021}. An even lower occurrence was found by \citet{Bryant2023}, who calculated an occurrence rate of $0.108\pm0.083$\% for short-period giant planets ($P < 10$~days; $R > 6.7~R_\oplus$) around stars with masses between 0.26 and 0.42~$M_\odot$ based on TESS transit detections. 

In addition to being a rare giant planet around a small star, TOI-3884\,b is also unusual in that its transits show persistent starspot crossings shortly following ingress and lasting about half the transit duration. The spot crossing signatures, which manifest as bumps in the transit light curves as the planet occults portions of the stellar disk that are cooler and fainter than the unspotted photosphere, are present in three sectors of Transiting Exoplanet Survey Satellite \citep[TESS;][]{Ri15} data, as well as multiple epochs from ground-based facilities (\citetalias{Almenara2022}; \citetalias{LibbyRoberts2023}). In total, a spot crossing event has been observed in 25 different transit observations of TOI-3884~b taken over a baseline of over two years.

As explained in \citetalias{LibbyRoberts2023}, there are two system configurations that could produce a spot crossing event in every transit observation. In the first scenario, the star has a single large spot (or multiple smaller spots) spanning a distance of about $0.44~R_\star$ across the face of the stellar surface. The orbital period of the planet  would have to be an integer multiple of the rotation period of the star, so that the spot is always in the same location when the transit is observed.

In the second scenario, the star is significantly inclined so that one of its poles is directed toward the observer, and the star has a large, long-lived spot (or spots) located at or near the pole. Large polar starspots are commonly observed on dwarf stars with the Doppler imaging technique \citep[e.g.,][]{Berdyugina1999, Strassmeier2002, Waite2015, Willamo2019, Cang2020}. They have generally been identified on active stars that rotate rapidly \cite[for dwarf stars, rotation periods less than about 5 days; see Table~2 of][]{Strassmeier2009}, but it should be noted that the sensitivity of Doppler imaging is biased towards stars with high $v\sin{i}$, for which rotational broadening dominates the stellar line widths \citep{Vogt1983}. It has been suggested that rapid rotation allows the Coriolis force to dominate over the buoyancy force, causing magnetic flux tubes to necessarily emerge at high latitudes \citep{Schuessler1992}. Alternatively, polar spots could form and be maintained through the transport of flux tubes in the meridional flow of rapidly rotating stars \citep{Schrijver2001, Mackay2004, Holzwarth2006}. If the nearly polar spot scenario is the reality for TOI-3884, its transiting planet would have to be in a misaligned orbit to exhibit a starspot crossing every transit. 

While the two scenarios could potentially be distinguished with a measurement of the stellar rotation period, TESS data showed no significant long-term variation that could be attributed to rotational modulation (\citetalias{LibbyRoberts2023}).
Spectroscopic measurements of the projected stellar velocity $v\sin{i}$ could help constrain the maximum rotation period given the stellar radius, but \citetalias{Almenara2022} and \citetalias{LibbyRoberts2023} reported very different values of $v\sin{i} = 1.1$~\kms and $3.6\pm0.9$~\kms respectively. \citetalias{LibbyRoberts2023} argued in favor of the polar spot explanation for TOI-3884 on the basis of the non-detection of rotation modulation: the first, aligned, scenario
would induce large (approximately $10\%$ peak-to-peak amplitude) photometric variations that were not observed in three sectors of TESS data. Their model of the misaligned orbit plus polar spot scenario could produce photometric modulation with an amplitude of $<0.5\%$, below the noise level of the TESS photometry, leading them to suggest that this was the more likely of the two scenarios. 
The polar starspot scenario was also explored recently by \citet{Sagynbayeva2025}, who applied a hierarchical Bayesian framework and probabilistic spot model to model the previously published TESS data.

In this paper, we present new photometric observations from the \textit{Tierras} Observatory that confirm that TOI-3884~b is in a misaligned orbit over a polar spot. We measure the stellar rotation period from both the long-term photometric modulation and changes in the planet's transit shape across multiple orbits. The transit shape variations also allow us to determine the relative alignment of the star's spin axis and the planet's orbit. 
We describe the observations in Section~\ref{sec:obs} and their analysis in Section~\ref{sec:analysis}, and discuss our results in Section~\ref{sec:discussion}.

\section{Observations} \label{sec:obs}
\subsection{Tierras}
Our dataset comprises primarily photometric observations from the \textit{Tierras} Observatory taken on 124 nights between UT 2024~November~13 and \tierrasenddate. \textit{Tierras} is located at Fred Lawrence Whipple Observatory (FLWO) on Mount Hopkins in Arizona. We gathered data with a 60-s exposure time in the custom \textit{Tierras} filter ($\lambda_C = 863.5$~nm, FWHM $=40$~nm) which was designed to limit photometric errors due to precipitable water vapor variability to less than 250 parts-per-million \citep{GaMe20}, enabling excellent night-to-night stability. On nights without transits, we performed between one and three 5-minute visits to the TOI-3884 field, depending on observability and scheduling constraints. On nights with transits, we performed staring observations with time coverage up to approximately two transit durations surrounding the predicted mid-transit time. Throughout this work, we will refer to the transits by their transit epoch, referenced to the first transit event observed in TESS Sector 46 ($T_0 = 2459556.51669$~BJD$_\mathrm{TDB}$; \citetalias{LibbyRoberts2023}). In total, we observed six transits with \textit{Tierras}: Epochs 247, 254, 258, 263, 272, and 276.

We reduced the \textit{Tierras} data using the custom facility pipeline \citep{Tamburo2025}. We performed aperture photometry for all sources in the \textit{Tierras}  images with Gaia $G_\mathrm{RP} < 17$~mag\footnote{The chosen $G_\mathrm{RP} = 17$~mag cutoff corresponds to a photometric precision of approximately 1\% in 60-s exposures.}, of which there were \ntierrassources~(including TOI-3884). We used circular aperture radii ranging from 5--20 pixels (in increments of 1 pixel) corresponding to an angular range of 2.2\arcsec--8.6\arcsec. Apertures were placed using the world coordinate systems (WCS) of the images and the Gaia DR3 coordinates and proper motions of sources propagated to the \textit{Tierras} epoch. We estimated the local background for each source in each image using a circular annulus with an inner radius of 25~pixels and an outer radius of 35~pixels. We applied a 2$\sigma$ clipping to the pixels in the annulus and took the background to be the mean of the remaining values. 

The target's background-subtracted flux was corrected using an artificial light curve (ALC) that is the weighted sum of the 74 reference stars in the field. The ALC weights are determined in an iterative process that is designed to de-weight especially noisy or variable reference stars \citep{Broeg2005, Murray2020, Tamburo2022a}.  The weights were determined using exposures that met the following set of quality criteria: median normalized flux of all sources $> 0.5$, full-width half-max (FWHM) seeing $< 4\arcsec$, pointing error in both the x and y axes $< 20$~pixels, and WCS solutions with RMS $< 0.215\arcsec$ (half a \textit{Tierras} pixel). The aperture that produced the minimum scatter was selected for further analysis, which in this case used a 7-pixel radius. 

We found that the average standard deviation of the out-of-transit data during the six \textit{Tierras} transit observations was higher than the median pipeline-calculated uncertainties by a factor of 1.655. We therefore inflated the calculated uncertainties of the \textit{Tierras} data by this factor. 

\subsection{FLWO 1.2-m and LCOGT SSO 1-m}
We observed two transits with the 1.2-m telescope at Fred Lawrence Whipple Observatory (FLWO) at Mt. Hopkins, Arizona in $g^\prime$-band using KeplerCam. We took observations with an exposure time of 60~s. The $4096\times4096$ Fairchild CCD 486 detector has an image scale of $0\farcs672$ per $2\times2$ binned pixel, resulting in a $23\farcm1\times23\farcm1$ field of view. Photometric data were extracted using {\tt AstroImageJ} \citep{Collins:2017} with circular apertures. The first of these, Epoch 258, was observed contemporaneously with \textit{Tierras}. The second, Epoch 267, was only partially observed due to poor weather conditions, resulting in coverage of the first half of the transit (and no observations with \textit{Tierras}, which has stricter opening conditions). 

We observed the second half of the transit on Epoch 267 with the Las Cumbres Observatory Global Telescope (LCOGT) \citep{Brown:2013} 1\,m network node at Siding Spring Observatory near Coonabarabran, Australia (SSO) using 120\,s exposures in Sloan $i'$ band. The LCOGT 1\,m telescopes are equipped with a $4096\times4096$ SINISTRO camera having an image scale of $0\farcs389$ per pixel, resulting in a $26\arcmin\times26\arcmin$ field of view. The images were calibrated by the standard LCOGT {\tt BANZAI} pipeline \citep{McCully:2018} and differential photometric data were extracted using {\tt AstroImageJ}. The stellar point-spread-functions (PSFs) in the images were not circular due to poor intra-exposure telescope guiding and tracking, likely caused by the high airmass (1.49-2.65) observations of the northern hemisphere target star from SSO. To optimize photometric precision, we used an elliptical photometric aperture with equivalent radius of 11 pixels ($4\farcs3$) and eight comparison stars. In each image, we fitted the aperture's eccentricity and angle to the PSF elongation of the target and eight comparison stars, while constraining each elliptical aperture to have the same number of pixels as a circular aperture with radius 11 pixels. The pixel counting accounts for the partial pixels along the boundary of the ellipse. 

Similar to the \textit{Tierras} observations, we found that the standard deviation of the  out-of-transit data for both the FLWO~1.2-m and LCO~SSO~1-m was greater than the reported pipeline uncertainties, by factors of 1.529 and 1.330, respectively. We inflated the uncertainties on these two data sets by these factors. 

We summarize the transit observations  in Table~\ref{tab:transits}. 

\begin{deluxetable*}{lrlrrrrl}[ht!]
\label{tab:transits}
\tablecaption{The transit observations of TOI-3884~b that we obtained in addition to sparse photometric monitoring with \textit{Tierras} from 2024~November~12 to \tierrasenddate.}
\tablehead{
    \colhead{UT Date} & 
    \colhead{$N$\tablenotemark{\scriptsize a}} & 
    \colhead{Facility} & 
    \colhead{Filter} & 
    \colhead{\makecell{Avg. Seeing \\ FWHM (\arcsec)}} & 
    \colhead{$\sigma$ (ppt)\tablenotemark{\scriptsize b}} & 
    \colhead{Airmass\tablenotemark{\scriptsize c}} &
    \colhead{Notes} 
}
\startdata
2025 Jan 03 & 247 & $Tierras$ & $Tierras$ & 2.16 & 6.1 & 1.20 $\rightarrow$ 1.06 $\rightarrow$ 1.08 & Observation ended at morning \\
 & & & & & & & nautical twilight\\
2025 Feb 04 & 254 & $Tierras$ & $Tierras$ & 2.16 & 5.3 & 2.25 $\rightarrow$ 1.06 $\rightarrow$ 1.08 & --- \\
2025 Feb 22 & 258 & $Tierras$ & $Tierras$ & 2.41 & 5.3 & 1.10 $\rightarrow$ 1.74 & Observation ended at morning \\
& & & &  & & & nautical twilight \\
2025 Feb 22 & 258 & FLWO 1.2-m & $g'$ & 5.10 & 9.4 & 1.06 $\rightarrow$ 1.66 & Observation ended at morning \\
& & & &  & & & nautical twilight \\
2025 Mar 17 & 263 & $Tierras$ & $Tierras$ & 1.91 & 10.1 & 1.70 $\rightarrow$ 1.07 & Cirrus clouds throughout \\
2025 Apr 04 & 267 &  FLWO 1.2-m & $g'$ & 2.58 & 6.4 & 1.23 $\rightarrow$ 1.54 & First half observed; significant \\
 &  &  & & &  & & correlation with airmass \\
2025 Apr 04 & 267 &  LCO SSO 1-m & $i'$ & 3.59 & 1.8 & 2.51 $\rightarrow$ 1.47 & Second half observed \\
2025 Apr 27 & 272 & $Tierras$ & $Tierras$ & 2.21 & 5.1 & 1.28 $\rightarrow$ 1.06 & Observation began at evening \\
 &  & &  & &  & & astronomical twilight \\
2025 May 15 & 276 & $Tierras$ & $Tierras$ & 3.09 & 5.3 & 1.17 $\rightarrow$ 2.21 & Significant correlation with \\
 &  &  &  & & &  & airmass; observation ended at\\
  &  &  &  & & &  & an hour angle of +4.25\\
\enddata
\tablenotetext{a}{Transit number $N$ is referenced with respect to the first transit of TOI-3884~b in TESS Sector 46.} 
\tablenotetext{b}{Standard deviation $\sigma$ is calculated using the out-of-transit data on the night in question.} 
\tablenotetext{c}{The first number in each row gives the starting airmass and the last number gives the ending airmass. If applicable, the minimum airmass is also indicated as an intermediate value.}
\end{deluxetable*}

\section{Analysis} \label{sec:analysis}

\subsection{Photometric Rotation Period}\label{subsec:rotation_period}

\begin{figure*}[ht!]
    \centering
    \includegraphics[width=0.9\textwidth]{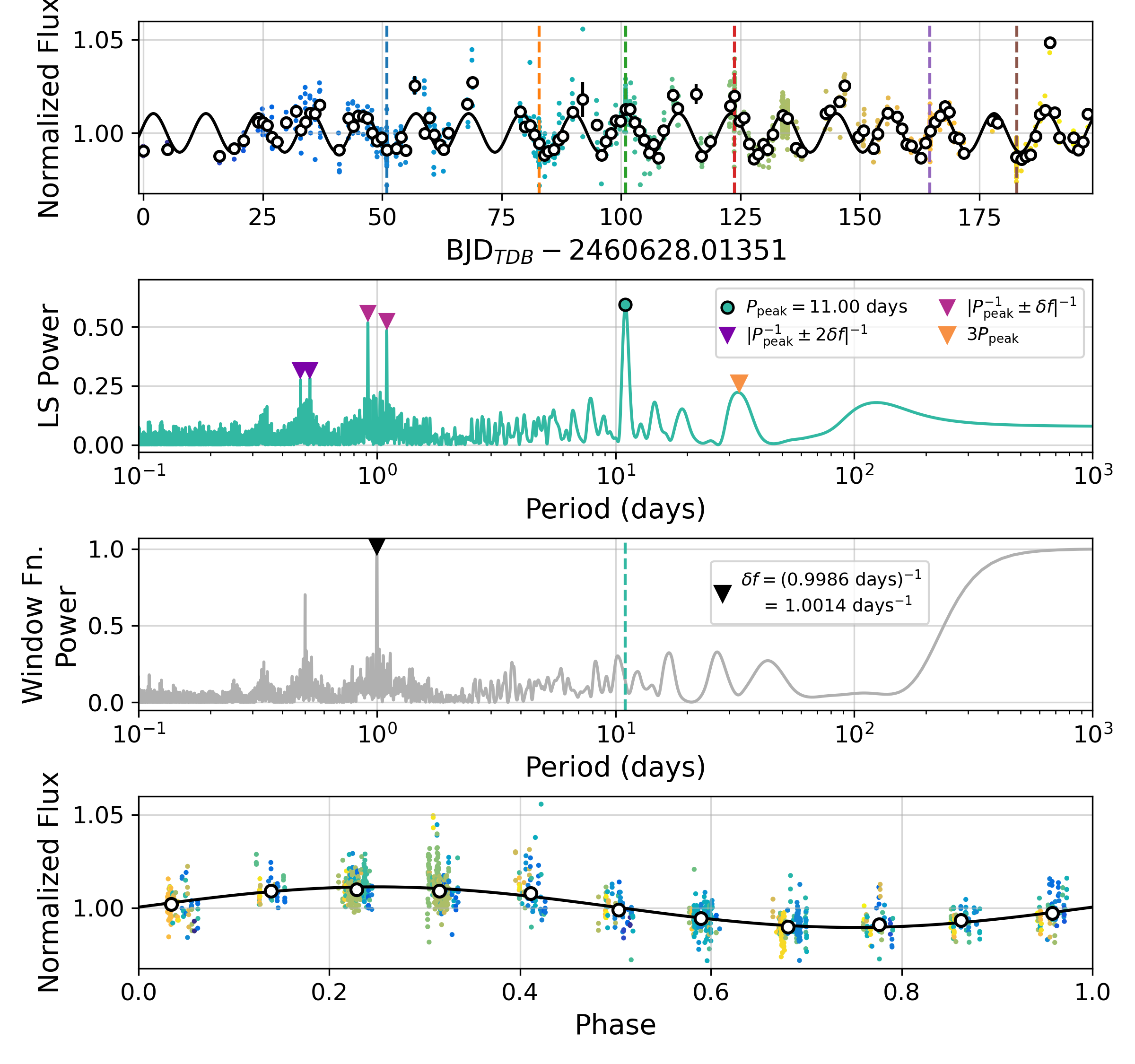}
    \caption{First panel: The \textit{Tierras} light curve of TOI-3884. Individual 60-s exposures are shown as points color-coded by time, while the medians fluxes measured on each night are shown as white points with error bars. Dashed lines indicate the epochs of the transit observations summarized in Table~\ref{tab:transits}, which have been masked out for the periodogram analysis. These data have been corrected for a long-term linear slope (see Section~\ref{subsec:starspot_modeling}), and a sine model (black) has been fit to the data as a visual aid. Second panel: The Lomb-Scargle periodogram of the data. There is a significant peak in the periodogram at \tierraslsper~days. Triangles indicate the locations of various aliases of the 11-day peak. Third panel: The window function of the data. A vertical dashed blue line indicates the location of the peak of the periodogram in the second panel. A black triangle indicates the peak of the window function, which occurs at approximately 1 day. Fourth panel: The \textit{Tierras} data phase-folded on a period of \tierraslsper~days. The colors of the points correspond to those in the top panel, and are used to indicate the progression of time. The black line shows a sine model fit to the phased data.}
    \label{fig:tierras_rotational_curve}
\end{figure*}

We show the \textit{Tierras}  sparse photometric monitoring light curve (i.e., with transits masked out) in Figure~\ref{fig:tierras_rotational_curve}. These data have also been corrected for a long-term linear slope that was present in the data (see Section~\ref{subsec:starspot_modeling}). A Lomb-Scargle (LS) periodogram of the data reveals a highly significant single peak at $P_\mathrm{peak} = $~\tierraslsper~days. Following \citet{Vanderplas2018}, we calculated the window function of the data (shown in the third row of Figure~\ref{fig:tierras_rotational_curve}) and searched for aliases of the 11-day signal. We identified the peak of the window function at $0.9986$~days, giving $\delta f = 1/0.9986~\mathrm{days}^{-1} = 1.0014~\mathrm{days}^{-1}$. This explains the peaks in the periodogram at $|P_\mathrm{peak}^{-1} \pm \delta f|^{-1}$ and $|P_\mathrm{peak}^{-1} \pm 2\delta f|^{-1}$, which we indicate with pink and purple triangles, respectively. We found another significant peak at $3P$, but when folding the data on this period, a clear sinusoidal signal remains. We therefore conclude that the true peak of the periodogram, and thus the rotational period of TOI-3884, is indeed 11~days.  


\subsection{Starspot Modeling}\label{subsec:starspot_modeling}
We fit all the in-transit and out-of-transit data from the \textit{Tierras}, FLWO~1.2-m, and LSO~SSO~1-m telescopes using the \texttt{starry} Python package, which models the stellar surface flux using spherical harmonics \citep{Luger2019}. We used an $l=10$-degree spherical harmonic expansion to model the surface. We modeled a single top-hat spot on the stellar surface which is described with a radius $r_\mathrm{spot}$, contrast $c_\mathrm{spot}$, latitude $\phi_\mathrm{spot}$, longitude $\lambda_\mathrm{spot}$, and a Gaussian smoothing factor that is applied to the top-hat model, $s_\mathrm{spot}$. In our modeling, $c_\mathrm{spot}$ is not a free parameter; rather, we calculate its value in the three filters (\textit{Tierras}, $g^\prime$, and $i^\prime$)  using sampled stellar and spot effective temperatures, $T_\mathrm{eff, \star}$ and $T_\mathrm{eff, spot}$, and SPHINX M-dwarf spectral models \citep{Iyer2023}. We first draw a random $T_\mathrm{eff, \star}$ and $T_\mathrm{eff, spot}$ from the prior distribution. We linearly interpolate the neighboring SPHINX models for a $\log{g} = 5.0$ and $Z=0.0$ source (\citetalias{LibbyRoberts2023}) for both the stellar photosphere and the spot to get an approximate spectral model at the drawn temperatures. Then, the contrast in a given filter is given by 

\begin{equation}
    c = 1-\frac{\int_{\lambda_1}^{\lambda_2}f_\mathrm{spot} t~d\lambda}{\int_{\lambda_1}^{\lambda_2}f_\star t~d\lambda} \,,
\end{equation}
where $f_\mathrm{spot}$ is the interpolated spot spectrum at temperature $T_\mathrm{spot}$, $f_\mathrm{\star}$ is the interpolated stellar photosphere spectrum at temperature $T_\mathrm{\star}$, $t$ is the filter transmission curve, and $\lambda_{1}$ and $\lambda_2$ represent the lower and upper wavelengths of the filter transmission curve, respectively.

With the default spot smoothing level of $2/l=0.2$, \texttt{starry} can accurately model spots with radii as small as about 22.5~degrees\footnote{\url{https://starry.readthedocs.io/en/latest/notebooks/StarSpots/}}, which we take as a lower limit to the spot radii allowed in our modeling. We also fit for the rotation period of the star $P_\mathrm{rot}$, the stellar inclination $i_\mathrm{\star}$, stellar mass $M_\mathrm{\star}$, and the stellar radius $R_\mathrm{\star}$. We applied a quadratic limb-darkening law to the stellar surface in the three filters that comprise our observations. We placed a broad Gaussian prior on each limb-darkening coefficient with a standard deviation of 0.2. The central values for the coefficients in the \textit{Tierras} filter were taken to be  $u_{1,Tierras}=0.224$ and $u_{2,Tierras}=0.122$, the central values found in analysis of TESS data of TOI-3884 by \citet{Almenara2022}. We make this assumption because the TESS and \textit{Tierras} filters have similar effective wavelengths ($800$~nm and $863.5$~nm, respectively). Additionally, despite the very different filter widths ($401$~nm FWHM vs. $40$~nm FWHM, respectively), we find that the spot contrast is nearly identical when integrated across the TESS and \textit{Tierras} filters (see Table~\ref{tab:results}). Regardless, the width of 0.2 on both coefficients allows for significantly different limb darkening coefficients to be sampled if they are supported by the \textit{Tierras} data. For the $g^\prime$- and $i^\prime$-bands, we took the central values corresponding to a $T_{\mathrm{eff}}=3200$~K, $\log{g} = 5.0$, $Z=0.0$ star from \citet{Claret2011}, with $u_{1, \mathrm{g^\prime}}=0.4763$, $u_{2, g^\prime} =0.4000$, $u_{1, \mathrm{i^\prime}} = 0.2800$, and $u_{2, \mathrm{i^\prime}} = 0.3293$.

We placed a planet in a Keplerian orbit around this spotted host star. We fixed the planetary mass to 32.59~$M_\oplus$, the orbital period to 4.5445828~days, and the transit mid-time at Epoch 0, $T_0$, to $2459556.51669$ (\citetalias{LibbyRoberts2023}). We fit for the stellar sky-projected obliquity $\lambda_\star$, the scaled radius of the planet $R_\mathrm{p}/R_\star$, the inclination of the planetary orbit $i_\mathrm{p}$, the eccentricity of the orbit $e$, and the longitude of periastron $\omega$. We note that in \texttt{starry}, $\lambda_\star$ is defined as the angle of rotation on the sky measured counter-clockwise from north\footnote{\url{https://starry.readthedocs.io/en/latest/notebooks/Orientation/}}, which in the coordinate system of a planet orbiting left to right along the $x$ axis that we use throughout this work, corresponds to the orbit normal. This is different from the conventional definition of $\lambda_\star$ in the literature, which is measured clockwise from the sky projection of the orbit normal, $n_\mathrm{orb}$ (north in our coordinate system) to the sky projection of the stellar angular momentum vector, $n_\mathrm{*}$ \citep[e.g.,][]{Albrecht2022}. Throughout this work, we report $\lambda_\star$ using the conventional definition.

We also fit for a number of nuisance parameters. We found evidence for a long-term linear decrease in the flux in the \textit{Tierras} data, and correct for this by fitting for a global slope $m_{\mathrm{global},~Tierras}$ and offset $b_{\mathrm{global},~Tierras}$. To test whether similar long-term trends were present in the comparison star light curves, we fit each with a linear function. We found that the long term slope we fit for TOI-3884 ($-6.3e-5\pm5.1e-6$~days$^{-1}$; see Table~\ref{tab:results}) was comparable to the distribution of long-term slopes measured for the 74 reference stars, which had a median value of $1.8e-5^{+5.6e-5}_{-5.5e-5}$~days$^{-1}$ (16\textendash 84 percentile range). We therefore suspect that long-term decrease in brightness observed for TOI-3884 is non-physical, but is rather a systematic effect in our data. Regardless of its origin, our modeling of the linear slope ensures that it does not affect our results for the rotation period and transit modeling.

We also allowed for offsets on the individual transit nights in the \textit{Tierras} data (adding six parameters), as night-to-night uncertainties in the flux level could bias the model if unaccounted for. We found that residuals from Epoch 276 were significantly correlated with airmass, and included a linear airmass correction term for this transit only in the \textit{Tierras} data. Finally, we included an additional nightly offset for the night of UT~2025~May~22, which were taken during an apparent flare event. TOI-3884 is an active star with an H$\alpha$ equivalent width of $-3.86\pm0.02$~\AA, and flare events have been previously observed in its TESS data (\citetalias{LibbyRoberts2023}). 

For both the FLWO~1.2-m and LCO~SSO~1-m, we normalized each transit using the out-of-transit baseline of the observations and fit for a flux offset in the model. We also allowed for a linear airmass correction for each of these transits. 

\startlongtable
\begin{deluxetable*}{llLR}
\tabletypesize{\small}
\label{tab:results}
\tablecaption{Priors and Fit Results for Starspot Analysis}
\tablehead{\colhead{Parameter} & 
\colhead{Description} & 
\colhead{Prior} & \colhead{Posterior\tablenotemark{\scriptsize a}}}
\input{fit_results}
\tablenotetext{a}{Median and $16-84$ percentile range}
\tablenotetext{b}{\citet{LibbyRoberts2023}}
\tablenotemark{c}{Note that our analysis is more sensitive to the difference between the stellar and spot effective temperatures, $\Delta T$, than to the absolute temperatures $T_\mathrm{eff, \star}$ and $T_\mathrm{eff, spot}$. We thus caution against using our posterior value for $T_\mathrm{eff,\star}$ as the true temperature of TOI-3884's photosphere, and instead defer to the measurement reported from SED fitting in \citetalias{LibbyRoberts2023}.} 
\end{deluxetable*}

\begin{figure*}
    \centering
    \includegraphics[width=\textwidth]{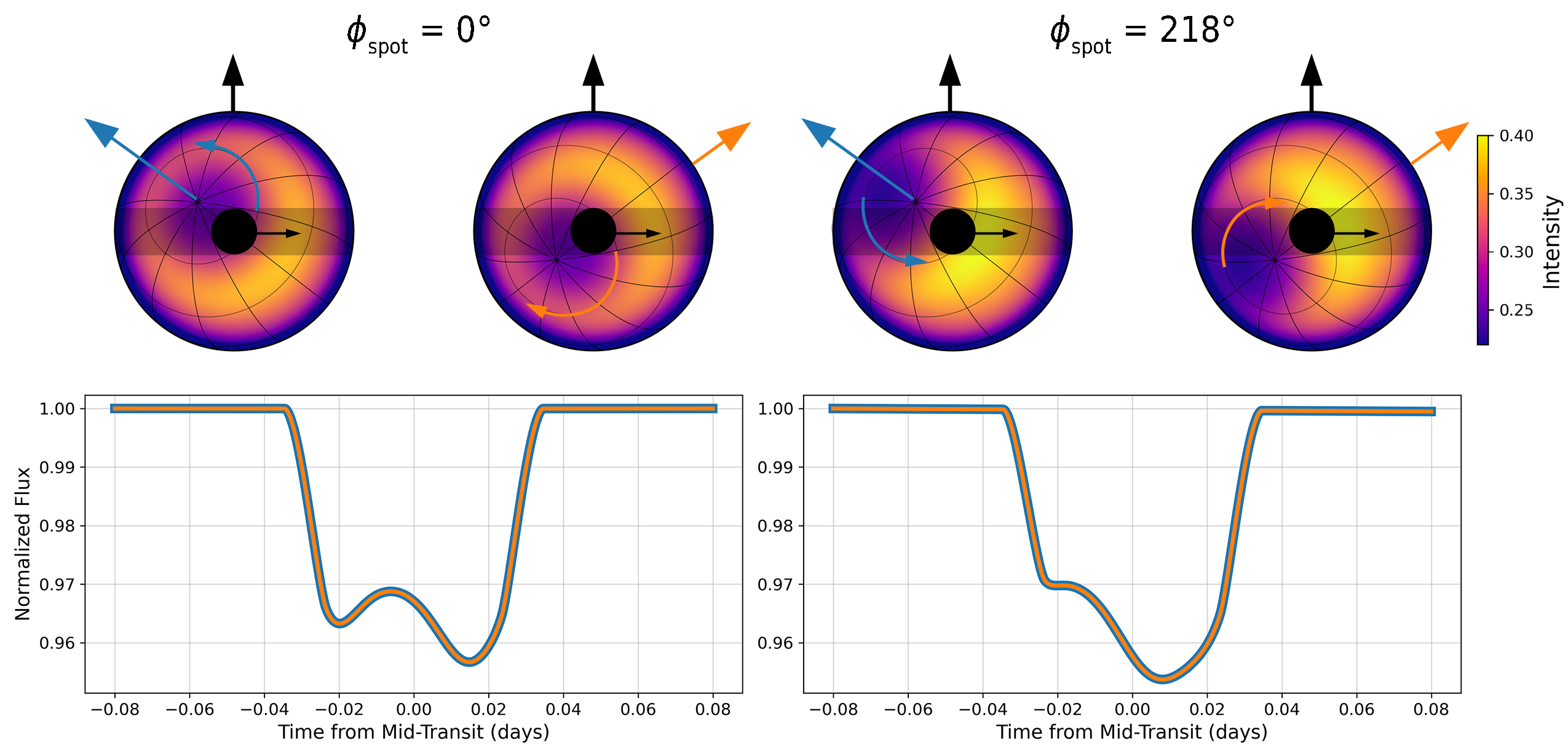}
    \caption{An illustration of the degeneracy between ``positive" pole solutions and ``negative" pole solutions during two different transit events: one with the spot located at longitude $\phi_\mathrm{spot} = 0^\circ$, the other at $\phi_\mathrm{spot} = 218^\circ$}. For each event, a positive surface configuration is shown (blue) with $i_\star = 22.3^\circ$, $\lambda_\star = 305.1^\circ$, $i_\mathrm{p} = 89.934^\circ$, and and $\lambda_\mathrm{spot} = 80.5^\circ$, along with the conjugate negative solution (orange):  $i_\star^\prime = 180^\circ - 22.3^\circ = 157.7^\circ$, $\lambda_\star^\prime = 360^\circ - 305.1^\circ = 54.9^\circ$, $i_\mathrm{p}^\prime = 180^\circ - 89.934^\circ = 90.066^\circ$, and $\lambda_\mathrm{spot}^\prime = -80.5^\circ$. The planet's transit chord is indicated with a shaded rectangle and its orbital direction is indicated with an arrow. The planet's orbital angular momentum vector is indicated with a vertical black arrow. A blue (orange) arrow indicates the direction of stellar spin rotational axis for the positive (negative) configuration. A curved arrow indicates the direction of the spin for the pole in each case. In the bottom panels, we show the light curve predicted for the positive and negative surface configurations as thick blue and thin orange lines, respectively. The light curves are exactly equivalent.
    \label{fig:degenerate_solutions}
\end{figure*}

In total, our model used 37 free parameters. We summarize the priors we placed on these parameters in Table~\ref{tab:results}. 
Note that we placed a uniform prior on $i_\star$ between $0^\circ$ and $90^\circ$. With the convention used in \texttt{starry}, this only permits solutions that present the ``positive" rotational pole to the observer, meaning the pole that rotates in a right-handed (or counter-clockwise) sense by IAU definition \citep{Archinal2011}.
There is an additional family of solutions for which $90^\circ < i_\star < 180^\circ$, where the ``negative" rotational pole (which rotates clockwise) is visible to the observer. However, solutions in this family result in identical photometry to the first case, by transforming $i_\star^\prime = 180^\circ - i_\star$, preserving the true stellar obliquity $\psi_\star$, and reflecting the spot configuration and transit chord about a plane parallel to the planet's orbital plane and passing through the center of the star. 
We illustrate this symmetry in Figure~\ref{fig:degenerate_solutions}, which shows the surface configuration for conjugate positive and negative pole configurations, along with the resulting light curves during two different transit events. The positive configuration was generated with $i_\star = 22.3^\circ$, $\lambda_\star = 305.1^\circ$, $i_\mathrm{p} = 89.934^\circ$, and $\lambda_\mathrm{spot} = 80.5^\circ$; the negative configuration was generated with $i_\star^\prime = 180^\circ - 22.3^\circ = 157.7^\circ$, $\lambda_\star^\prime = 360^\circ - 305.1^\circ = 54.9^\circ$, $i_\mathrm{p}^\prime = 180^\circ - 89.957^\circ = 90.066^\circ$, and $\lambda_\mathrm{spot}^\prime = -80.5^\circ$.
While these two configurations correspond to different physical scenarios, the symmetry of the system means we cannot determine whether the planet is transiting over the positive or negative rotational pole with photometric data alone.
Given this degeneracy, we elected to limit our analysis to solutions about the positive pole, motivating the uniform prior on $i_\star$ between $0^\circ$ and $90^\circ$ given in Table~\ref{tab:results}.

{\subsection{Results}
\label{subsec:results}
We ran 62 walkers for $100{,}000$ steps using the Python package \texttt{emcee} \citep{Fo13} and discarded the first $10{,}000$ steps to account for burn-in. We report the median and 68\% confidence interval for the model free parameters in Table~\ref{tab:results}. The best-fit global model had a $\chi^2$ value of $1869.8$. With $2019$ data points and 37 free parameters, this gives a reduced $\chi^2$ value of 0.94, indicating that our model is a good fit to the data.

\begin{figure*}[ht!]
    \centering
    \includegraphics[width=\textwidth]{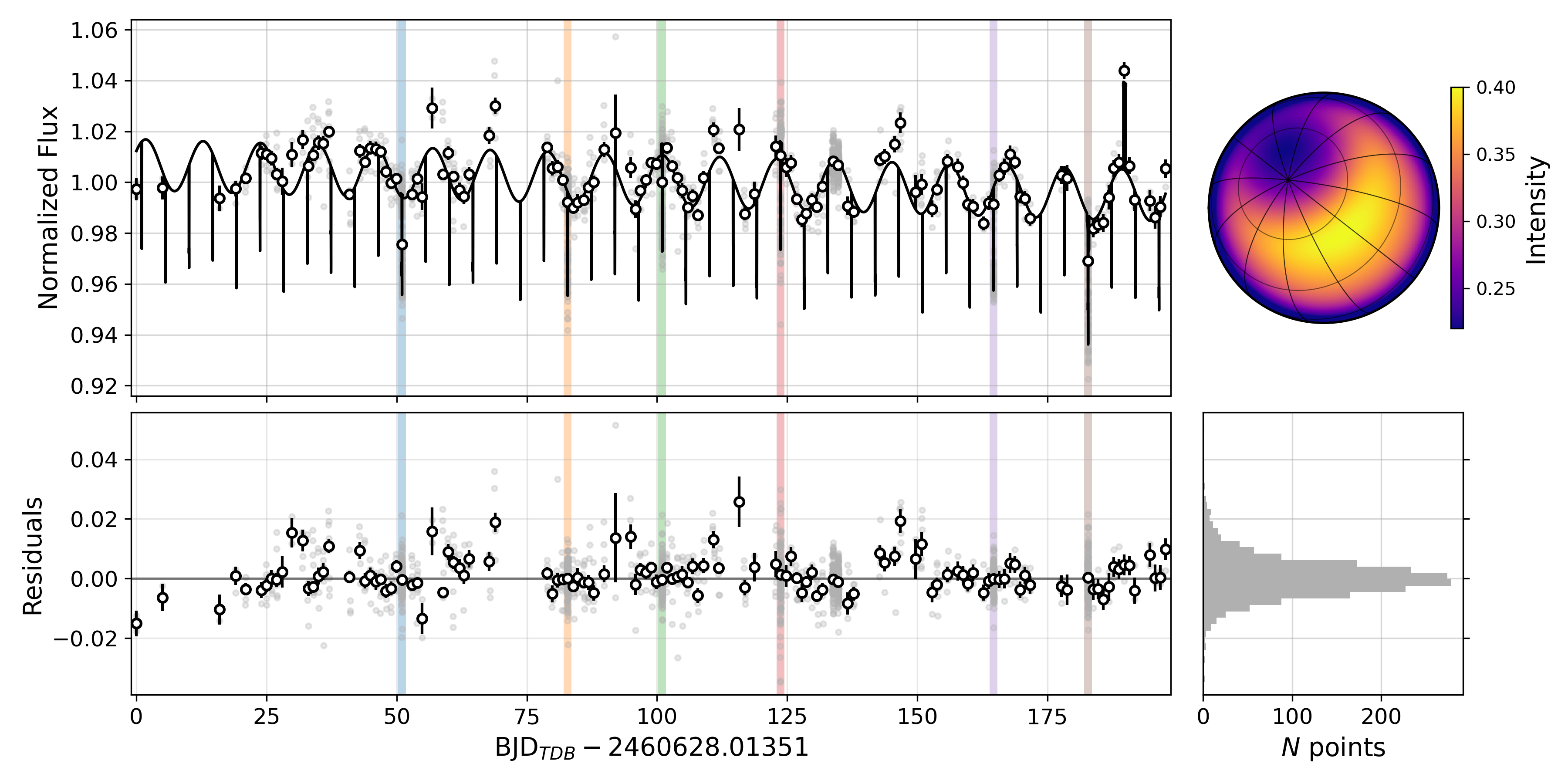}
    \caption{Top left: The full \textit{Tierras} light curve at native cadence (gray points) and binned over each night (white points with black outlines) along with the best-fit MCMC model (black). Colored regions indicate the nights during which transit observations occurred. Bottom left: The residuals from the best-fit model. Top right: The best-fit \texttt{starry} map at the first \textit{Tierras} time stamp. By construction, our models includes solutions for the positive stellar rotational pole only, meaning the spot is rotating counter-clockwise. Bottom right: A histogram of the residuals of the data. The residuals have a standard deviation of 6.9~ppt, while the residuals binned over each individual night have a standard deviation of 6.2~ppt.}
    
    \label{fig:full_model}
\end{figure*}

We show the full \textit{Tierras} light curve of TOI-3884 in the top left panel of  Figure~\ref{fig:full_model} along with the best-fit model from the MCMC analysis. The nights of the six \textit{Tierras} transit observations (see Table~\ref{tab:transits}) are indicated with shaded regions. The top-right panel shows the stellar surface corresponding to the best-fit model at the time of the first \textit{Tierras} exposure. It consists of a highly inclined star ($i_\star = $~\tierrasmcmcistar) that is significantly misaligned with the planet's orbit ($\lambda_\star = $~\tierrasmcmclambda). The star has a large spot ($r_\mathrm{spot}=$~\tierrasmcmcrspot ) located at \tierrasmcmclatspot~latitude.

As the spot travels around with the star's \tierrasmcmcprot-day rotation period, it alternates between being fully and partially visible to the observer. This time-variable visibility causes the 1\%-amplitude sinusoidal flux variations that are clearly visible in the \textit{Tierras} photometry. The bottom left panel shows the residuals from the best-fit model. The unbinned residuals have a standard deviation of 6.9~parts-per-thousand (ppt), while the residuals binned over the duration of each night have a standard deviation of 6.2~ppt. The bottom right panel shows the distribution of the unbinned residuals, which is approximately Gaussian.

In Figure~\ref{fig:transits_and_configs}, we show the \textit{Tierras} transit observations in more detail. The left column gives the rotational phase of the spot at mid-transit. The middle column shows the six transit observations along with the best-fit model. Models generated from random MCMC samples are shown in magenta. For the sake of comparing the transit shapes, the data and the models have been normalized to the out-of-transit baseline of the best model at each epoch. The right column shows the system configuration at mid-transit for each observation. 

The changing location of the spot on the stellar surface during the various transit observations is clearly manifested in the differences in the shape of the in-transit data. For example, Transits 247, 254, and 276 were all taken during phases where the spot was almost maximally in the path of the transit chord, and all three show a slight increase in flux shortly following second contact, with the maximum occurring near mid-transit.
Transits 258 and 263 were both taken near a phase where the spot was less in the transit path and show correspondingly weaker spot-crossing signals, with the maximum of the spot crossing signal occurring closer to the start of the transits.
However, note that Transit 258 shows additional bumps that are not captured by our model and are perhaps indicative of additional spot crossings, which we discuss further in the following paragraph along with the FLWO~1.2-m data. Finally, Transit 272 was taken at a phase where the spot was almost entirely out of the transit chord and, as a result, is comparatively flat throughout the entire transit. This surface configuration could explain previous transit observations in which no clear spot crossings were observed
(\citetalias{Almenara2022}, \citetalias{LibbyRoberts2023}). 

\begin{figure*}
    \centering
    \includegraphics[width=\textwidth]{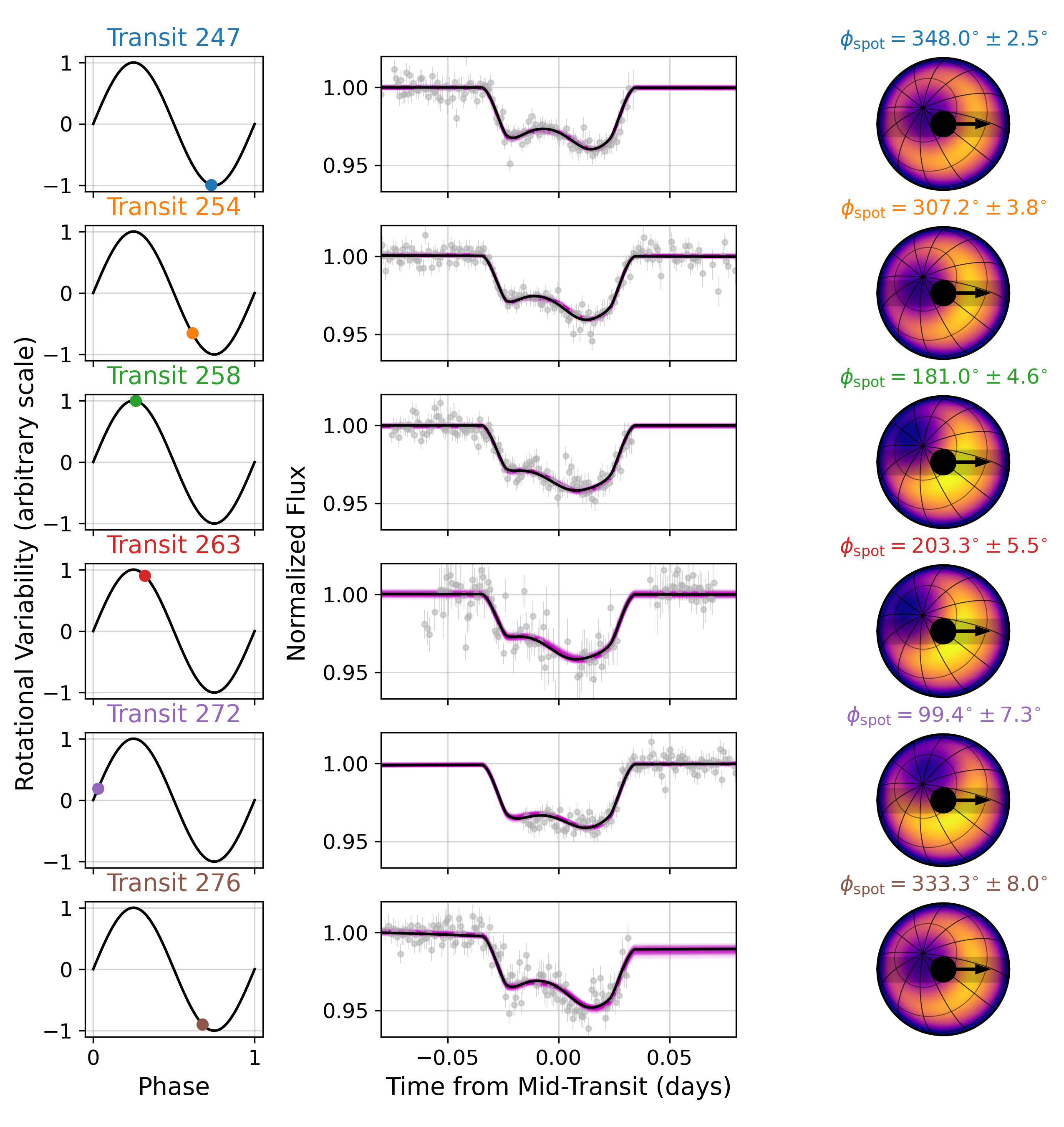}
    \caption{\textit{Tierras} transit observations of TOI-3884. The left column indicates the rotational phase of the star at mid-transit. The middle column shows the data and the best-fit transit model in black. The magenta lines show 100 models generated from random samples from the MCMC. For comparison, each observation has been normalized using the out-of-transit data on the given night. The models for Transit 276 show a slight slope due to an airmass correction that was applied to this event. The right column shows the configuration of the star and planet at mid-transit, with the spot contrast and limb darkening coefficients set to their best-fit values in the \textit{Tierras} filter. An arrow and shaded region indicate the path of the planet on its misaligned orbit across the stellar disk. The color scale of the stellar surface matches that in Figures~\ref{fig:degenerate_solutions} and \ref{fig:full_model}. The longitude of the spot at mid-transit is reported above each surface plot.}
    \label{fig:transits_and_configs}
\end{figure*}

In Figure~\ref{fig:transit_258}, we show detailed observations of Transit 258. This event was observed simultaneously by \textit{Tierras} and the FLWO~1.2-m telescope (in the \textit{Tierras} filter and $g^\prime$ band, respectively). The model favored a significant airmass coefficient for the FLWO~1.2-m observation of this transit, with a value of $0.0376^{+0.0060}_{-0.0053}$, resulting in models that have an appreciable curvature. Despite this adjustment, the models are a poor fit to both the \textit{Tierras} and FLWO~1.2-m data for this transit. In particular, we note that there is a feature following ingress that is not captured by the model in either filter.  We posit that this is due to an additional spot-crossing event in this light curve, which we are unable to fit with a single-spot model. A large main spot and multiple smaller spots were inferred by \citetalias{LibbyRoberts2023} to explain their transit observations, and this transit may provide further support to that possibility. To test this, we ran a second MCMC where we fit Transit 258 with a two-spot model. We locked all the stellar and planetary parameters to their values in the best-fit model from the one-spot MCMC run. To permit the accurate modeling of smaller spots, we ran the models with a spherical harmonic degree of $l=25$. We then modeled two spots on the stellar surface, and proceeded with our modeling as with the single-spot model. We show the best-fit two-spot model with a dashed line in Figure~\ref{fig:transit_258}, which places a smaller spot ($r_\mathrm{spot, 2} = {10.7^\circ}^{+3.0^\circ}_{-3.3^\circ}$) in the transit path. This model achieves a better fit to the data, especially in the FLWO~1.2-m observations. We find that for the \textit{Tierras} observations of Transit 258, the best-fit two-spot model produces residuals with a standard deviation 4\% lower than the best-fit single-spot model. For the FLWO~1.2-m observations of this event, the standard deviation of the residuals from the two-spot model are 11\% lower than the one-spot model. Thus, it is possible that during Transit 258, a smaller spot complex was in the transit chord that was not seen in any of the other transits. More transits taken at different stellar rotation phases will have to be observed to determine whether this is truly the case (assuming this smaller complex persists over long timescales). 

\begin{figure}
    \centering
    \includegraphics[width=\columnwidth]{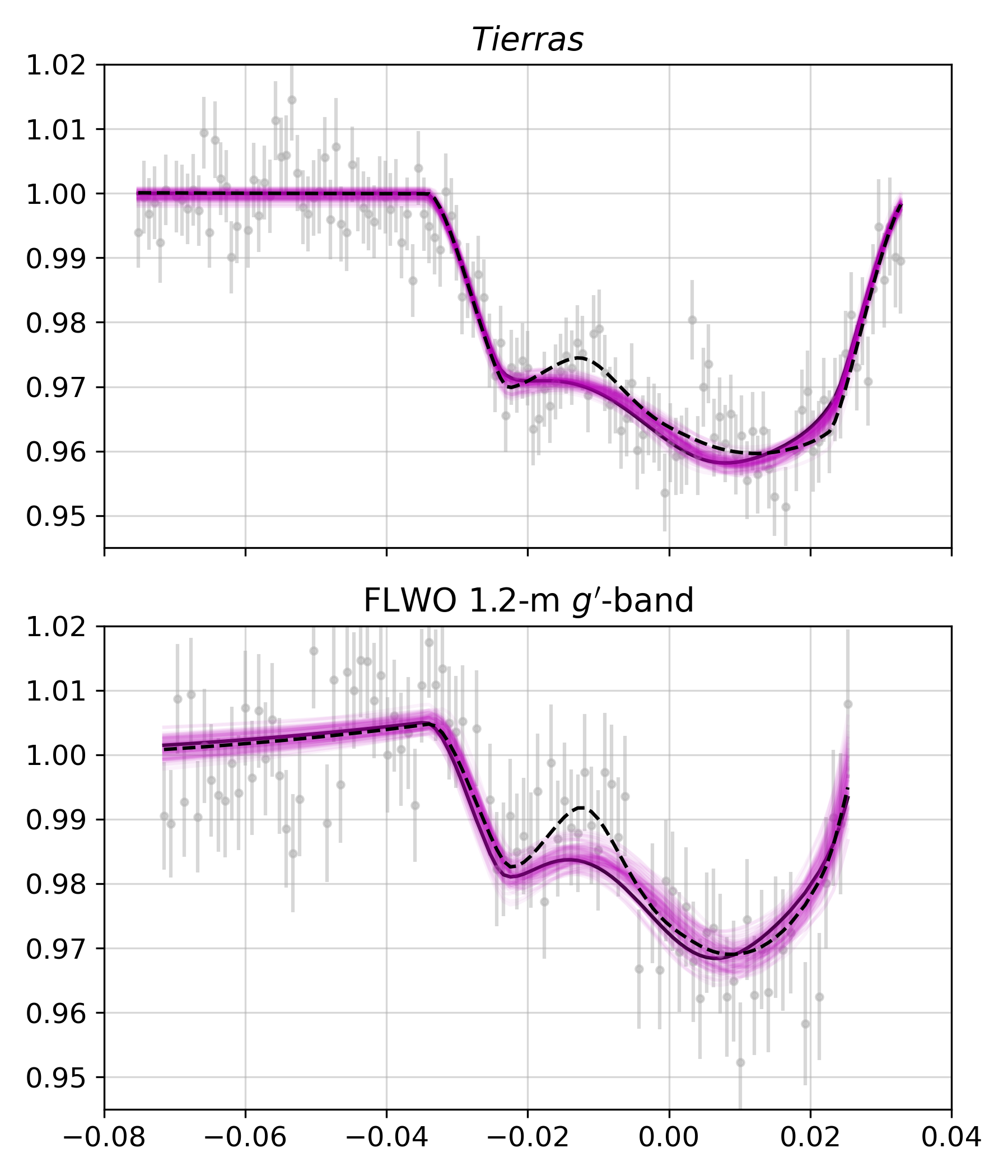}
    \caption{Simultaneous observations of Transit 258 with \textit{Tierras} (top panel) and FLWO~1.2-m in $g^\prime$-band (bottom panel). The best-fit model is shown as a solid black line and 100 models generated from random samples from the MCMC are shown in magenta. We have applied a linear airmass correction to the models fit to the FLWO~1.2-m data (see Table~\ref{tab:results}), resulting in models with appreciable curvature. A dashed black line shows the best-fit two-spot model, which achieves a better fit for this observation.}
    \label{fig:transit_258}
\end{figure}

In Figure~\ref{fig:transit_267}, we show the observations of Transit 267, which was partially observed by both the FLWO~1.2-m telescope in $g^\prime$-band and the LCO~SSO~1-m telescope in $i^\prime$-band. Between the two data sets, the whole transit was observed. These data are well fit by the single-spot model.

\begin{figure}
    \centering
    \includegraphics[width=\columnwidth]{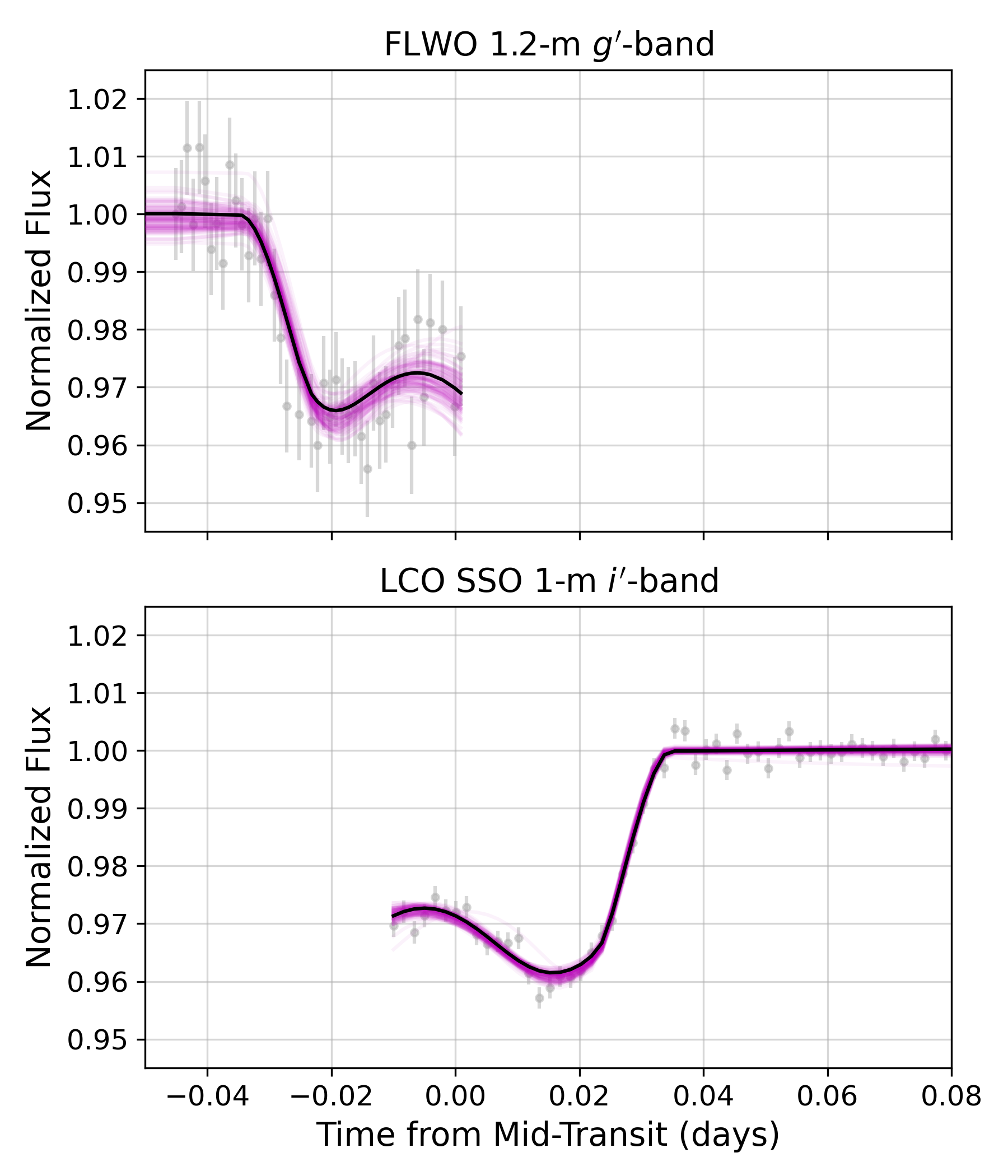}
    \caption{Simultaneous observations of Transit 267 with FLWO~1.2-m in $g^\prime$-band (top panel) and LCO SSO in $i^\prime$-band (bottom panel). The best fit model is shown in black and 100 models generated from random MCMC samples are shown in magenta.}
    \label{fig:transit_267}
\end{figure}

Finally, in Figure~\ref{fig:animation_keyframe}, we show the best-fit surface and the corresponding transit light curve model when the spot is located at longitude $\phi=0^\circ$. We provide an animation showing how the morphology of the spot-crossing event changes as a function of spot longitude in the HTML version of this article.


\begin{figure}
\begin{interactive}{animation}
{transit_shape_vs_spot_longitude.mp4}
\includegraphics[width=\columnwidth]{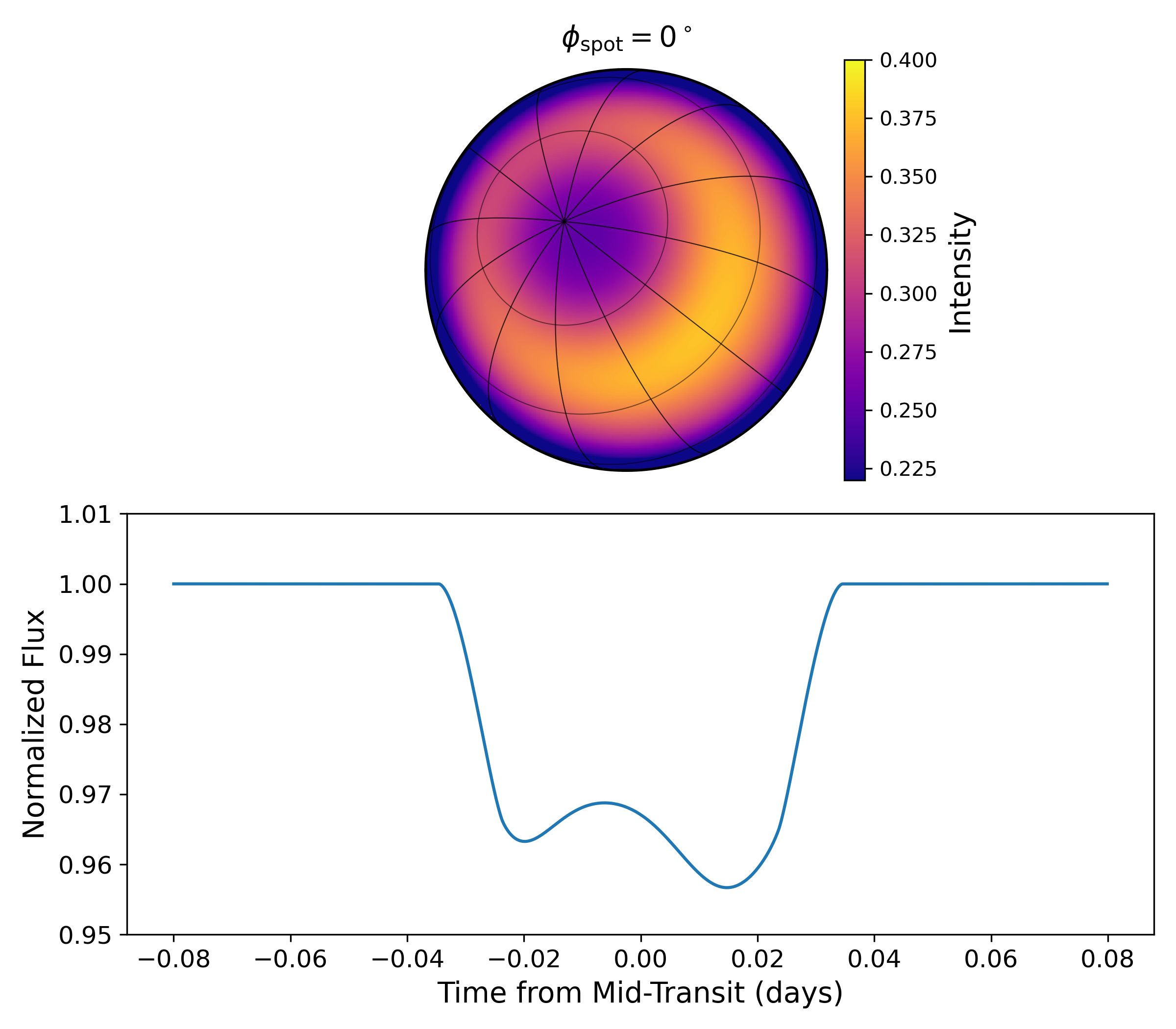}
\end{interactive}
\caption{The first frame of an animation showing how the stellar surface changes as the star rotates (top panel), and how the transit would look if it were observed with the corresponding surface configuration (bottom panel). The full animation, computed at $4^\circ$ intervals for one full rotation, is available in the HTML version of this article.}
\label{fig:animation_keyframe}
\end{figure}

\section{Discussion} \label{sec:discussion}

\subsection{The Geometry of the TOI-3884 System}
As discussed in Section~\ref{sec:intro}, \citetalias{LibbyRoberts2023} described two potential scenarios for the geometry of the TOI-3884 system that could result in persistent spot crossings. One of these possibilities is that the orbital period of the planet is an integer multiple of the rotation period of the star, such that every time the planet transits, the spot is in the same location from the perspective of the observer. While those authors disfavored this scenario on the basis that it would induce large photometric variations that were not seen in the TESS data, our results firmly rule it out. The measured stellar rotation period of \tierrasmcmcprot~days is not an integer fraction of the orbital period ($P_\mathrm{rot}/P_\mathrm{orb} = 2.4249\pm0.0033$), meaning the spot-crossing event occur at different phases across multiple transits. The only way to achieve the persistent starspot crossing events while simultaneously explaining the photometric variability in the \textit{Tierras} data is to have the planet on an orbit that takes it over the stellar pole, which hosts a large spot. This conclusion is also corroborated by the variations in the shape and timing of the spot-crossing event between transit epochs, which can be explained by a spot that is not perfectly centered at the pole and thus rotates in and out of the transit chord. These variations allow us to measure the stellar obliquity, and we find $\psi_\star = $~\tierrasmcmcpsi~degrees, indicating a significantly misaligned orbit for TOI-3884~b. 

\subsection{Archival Photometry}\label{subsec:archival_photometry}
We accessed archival \textit{g}-, \textit{r}-, and \textit{i}-band photometry of TOI-3884 from the Zwicky Transient Facility \citep[ZTF;][]{Bellm2018} in order to search for evidence of photometric variability prior to our observations. We retained the ZTF data that had not been flagged for poor quality and that had been taken at an airmass less than 1.3. This left us with 248 points in \textit{g}-band, 212 points in \textit{r}-band, and 81 points in \textit{i}-band. We also applied a median filter to the three light curves with a width of 180 days to remove long-term trends. We show the data in the top row of Figure~\ref{fig:ztf}, with time referenced to the time of the first \textit{Tierras} exposure. The $g$- and $r$-band light curves provide seasonal coverage starting about seven years before the \textit{Tierras} light curve, while the $i$-band light curve dates back about five years.

\begin{figure*}
    \centering
    \includegraphics[width=\textwidth]{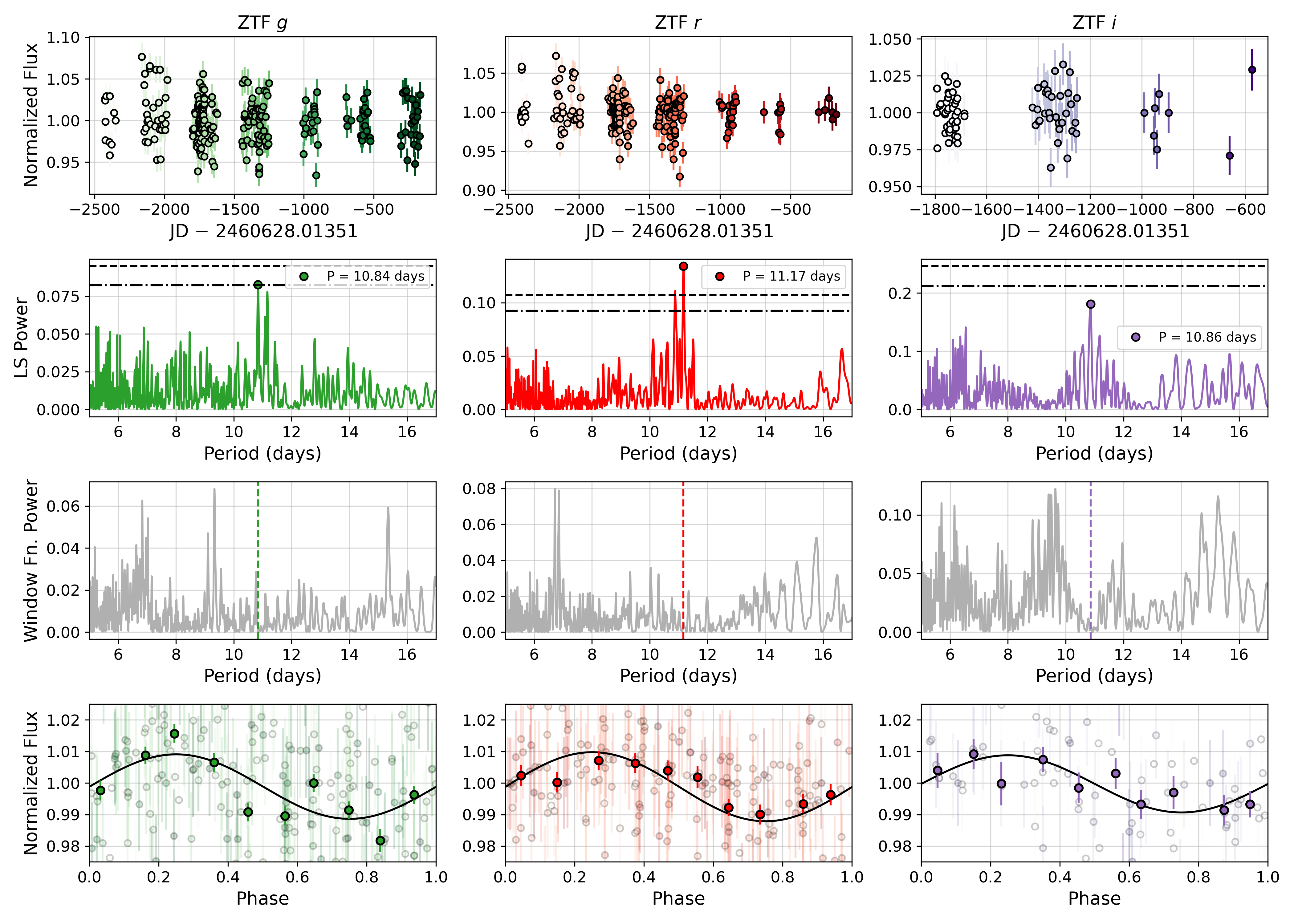}
    \caption{First row: Archival ZTF $g$-band (left column), $r$-band (middle column), and $i$-band (right column) photometry. Points are colored by time, which is calculated with respect to the first \textit{Tierras} exposure. Second row: The LS periodograms of the three light curves from 5--17 days, with the 5\% and 1\% FAP levels indicated with dash-dot and dashed lines, respectively. Each shows a peak around 11~days, though a peak with a FAP better than 1\% is only detected in the $r$-band data. Third row: The window functions of the data, with a dashed vertical line indicating the peak period of the periodograms in the second row. Fourth row: The light curves phase folded on the most significant periodogram peak. Binned data are shown as black-outlined points with error bars for clarity. A sine model fit is shown in each panel.}
    \label{fig:ztf}
\end{figure*}

We calculated the LS periodogram of each light curve and computed their window functions, which we show in rows two and three of Figure~\ref{fig:ztf}. All three light curves show a peak near 11~days in their LS periodograms, at 10.84~days in $g$-band, 11.17~days in $r$-band, and 10.86~days in $i$-band. We calculated the false-alarm probability (FAP) of each peak using the method of \citet{Baluev2008}, finding $FAP_\mathrm{g} = 4.7\%$, $FAP_\mathrm{r} = 0.046\%$, and $FAP_\mathrm{i} = 19\%$. Thus, while the detection is significant only in the $r$-band, the presence of a peak near 11-days in all three light curves suggests the possibility that the polar spot that we have inferred with the \textit{Tierras} data has persisted for at least seven years. Such long lifetimes are not uncommon for polar spots, as Doppler imaging observations of other dwarf stars have found evidence for polar spots across decades of observations \citep[e.g.,][]{Berdyugina1999}. In the bottom panel of Figure~\ref{fig:ztf}, we show the ZTF data phase-folded on the peak from the LS periodograms.

\subsection{Application to Previous Transit Observations}
TOI-3884 was observed by TESS in Sector 22 at 30-minute cadence and Sectors 46 and 49 at 2-minute cadence. Previous analyses averaged together the TESS transits for each sector when performing light curve fits (\citetalias{Almenara2022}, \citetalias{LibbyRoberts2023}). Here, we aim to determine whether there is evidence that the TESS transit light curves are modulated by the 11-day rotation period that we have identified in this work. 

We used \texttt{Lightkurve} \citep{Lightkurve2018} to access the Science Processing Operations Center \citep[SPOC;][]{SPOC2020} light curves for Sectors 46 and 49. We used the Pre-search Data Conditioning Simple Aperture Photometry (PDCSAP) flux and restricted the data to within one transit duration on either side of each transit time predicted from the linear ephemeris in \citetalias{LibbyRoberts2023}. Eight transits were observed in the two-minute cadence data, four in Sector 46 and four in Sector 49. We show the data in Figure~\ref{fig:tess_analysis}. 

\begin{figure}
    \centering
    \includegraphics[width=\columnwidth]{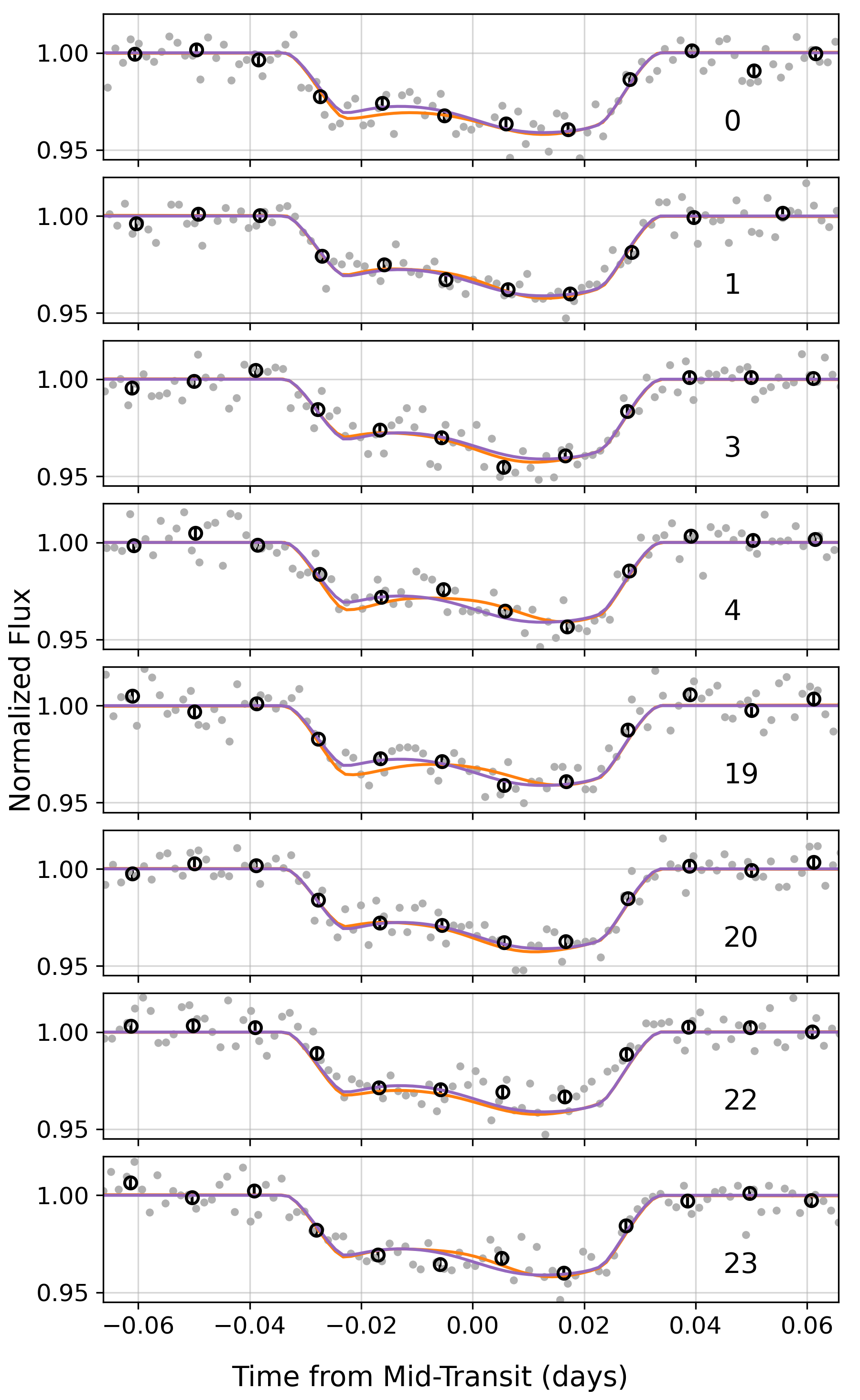}
    \caption{TESS transit observations of TOI-3884 from Sectors 46 and 49. The gray points are the two-minute cadence PDCSAP data from the SPOC analysis, while black points with error bars show the data binned over 16-minute intervals. The numbers in the lower right of each panel indicate the transit epoch referenced to the first transit in Sector~46. The best-fit spot model that includes an 11-day stellar rotation period is shown in orange, while the best-fit model with a static spot is shown in purple.}
    \label{fig:tess_analysis}
\end{figure}

We fit two sets of \texttt{starry} models to the TESS data, one where the rotation period was allowed to vary and one in which it was fixed to zero. In the first case, we placed a Gaussian prior on the rotation period using the posterior value given in Table~\ref{tab:results}. We fit for a single spot in both models, varying the radius, contrast, latitude, longitude, and smoothing, with priors given by the posteriors from Table~\ref{tab:results}. We also fit for quadratic limb darkening coefficients $u_1$ and $u_2$. We locked the stellar inclination and obliquity to their median posterior values in Table~\ref{tab:results}. We fixed $R_\mathrm{p}/R_\mathrm{\star}$ to $0.1808$, the median value from the TESS analysis of \citet{Almenara2022}. We ran 32 chains for $10{,}000$ steps for each model, and discarded the first $1{,}000$ steps for burn-in. 

The resulting best-fit models are shown in Figure~\ref{fig:tess_analysis}. For the model with stellar rotation, we found the best-fit model had a $\chi^2$ of $802.3$, while the corresponding value for the model without rotation was $785.9$. Thus, we find that the TESS data do not show evidence for variations in transit shape due to the spot rotation. This, combined with the non-detection of photometric variability in the TESS data, may suggest that the spot was located closer to the rotational pole during the TESS epochs. However, we caution that the lack of a detection of the 11-day rotation period in the TESS data is not necessarily surprising, as it has been shown that TESS data struggle to reliably detect such long rotation periods \citep[e.g., ][]{Holcomb2022, Boyle2025}. Additionally, the TESS data have significantly higher scatter than the \textit{Tierras} data. We measured the median standard deviation of the out-of-transit TESS data shown in Figure~\ref{fig:tess_analysis}, finding $\sigma=7.5$~ppt. We did the same thing for the \textit{Tierras} observations in Figure~\ref{fig:transits_and_configs} (binned to two minutes) and found $\sigma=4.0$~ppt. It is therefore possible that the TESS data simply lack the precision to detect the subtle spot-crossing variations seen in the \textit{Tierras} data.

\subsection{Spot Position During JWST Epochs}
\label{subsec:jwst}

TOI-3884 was selected for JWST Cycle 3 transmission spectroscopy observations in Guest Observer programs 5799 (22 hours; P.I. Garcia) and 5863 (24 hours; P.I. Murray). Observations for program 5863 were completed in December~2024, while those in program 5799 were scheduled to be performed in May and June~2025. As the interpretation of these data will likely be impacted by the precise location of the stellar spot, we provide the phase predicted by our model in Table~\ref{tab:jwst}. In this Table, the reported phase corresponds to the system used in Figure~\ref{fig:transits_and_configs}. We note that five out of the seven JWST transits fall within the time span of our \textit{Tierras} observations, which began on UT~13~November~2024 and ended on UT~30~May~2025. Given the stability of the rotational modulation over this time (see Figures~\ref{fig:tierras_rotational_curve} and \ref{fig:full_model}), we assume that the spot evolved (i.e., changed in position or size) negligibly between the end of our observations and the last two JWST transits, which were taken a little over two weeks after the end of the \textit{Tierras} light curve.

\begin{deluxetable*}{lrrr} \label{tab:jwst}
\tablecaption{JWST Cycle 3 transit observations of TOI-3884 and the predicted longitude of the starspot.}
\tablehead{\colhead{Time (UT)} & 
\colhead{$N$} & 
\colhead{Program ID} & 
\colhead{$\phi_\mathrm{spot}$ (deg.)}}
\startdata
16~Dec~2024~08:24 & 243 & 5863 & $109.8~\pm~4.9$ \\
20~Dec~2024~21:29 & 244 & 5863 & $258.3~\pm~5.0$ \\
25~Dec~2024~10:33 & 245 & 5863 & $46.7~\pm~5.1$ \\
29~Dec~2024~23:37 & 246 & 5863 & $195.1~\pm~5.1$ \\
24~May~2025~09:51 & 278 & 5799 & $265.3~\pm~9.1$ \\
02~Jun~2025~12:00 & 280 & 5799 & $202.2~\pm~9.4$ \\
16~Jun~2025~03:12 & 283 & 5799 & $287.5~\pm~9.9$ \\
\enddata
\end{deluxetable*} 

\subsection{Comparison with \citet{Mori2025}}
\label{subsec:mori}
As this manuscript was being prepared, \citet{Mori2025} published a similar analysis of the TOI-3884 system. Those authors obtained monitoring observations of TOI-3884 in $r$-band, finding strong evidence of 4.5\% peak-to-peak photometric variation at a rotation period of $11.043^{+0.054}_{-0.053}$~days. They also observed three transits of TOI-3884~b simultaneously in the $g$-, $r$-, $i$-, and $z$-bands with the MuSCAT3 and MuSCAT4 \citep{Narita2020} cameras on the 2m Faulkes Telescopes operated by LCO at Haleakala Observatory in Hawaii and Siding Springs Observatory in Australia, respectively. They modeled the transits with a single spot model using \texttt{fleck} \citep{Morris2020} and found evidence for a misaligned orbit ($\psi = {118.1^\circ}^{+5.6^\circ}_{-2.3^\circ}$) around a large ($r_\mathrm{spot} = 0.425 R_\star$) spot located near the stellar pole ($\lambda_\mathrm{spot} = {76.8^\circ}^{+4.6^\circ}_{-2.5^\circ}$). Qualitatively, our independent analysis confirms their result. However, we find significant differences between certain parameters in our model and those reported in \citet{Mori2025}, which we attempt to explain here. 

\citet{Mori2025} found a true obliquity $\psi_\star = 61.9^{\circ\,+2.3^\circ}_{\,-5.6^\circ}$. This is discrepant with the value of $\psi_\star$ from our modeling ($\psi_\star = $~\tierrasmcmcpsi) by 4.6$\sigma$. The difference can be explained by the higher impact parameter derived in \citet{Mori2025} compared to our work.
They reported that TOI-3884\,b transits its host star with an impact parameter of $b = 0.40\pm0.02$ and scaled semimajor axis $a/R_\star = 23.5\pm0.2$, substantially different from the corresponding parameters derived from our analysis of $b = 0.029^{+0.041}_{-0.021}$ and $a/R_\star = 25.06^{+0.91}_{-0.71}$.
Our results are more similar to the values of $b = -0.04\pm0.13$, $a/R_\star = 25.01\pm0.65$ and $b = 0.089^{+0.082}_{-0.061}$, $a/R_\star = 25.9^{+0.9}_{-0.7}$ reported by \citetalias{Almenara2022} and \citetalias{LibbyRoberts2023} respectively.
We suggest that the difference between our result and that of \citet{Mori2025} stems from a different parameterization of the transit model, where we placed priors on the physical mass and radius of the host star derived from the isochrone analysis performed by \citetalias{LibbyRoberts2023}, and computed the planet's transit parameters from those values.
We favor our solution of a nearly equatorial transit chord, as the high impact parameter reported by \citet{Mori2025} requires a stellar density of $\rho_\star = 11.85^{+0.30}_{-0.29}$\,\gcc to match the measured transit duration, lower than expected for a 0.3~$M_\odot$ star based on stellar models ($\rho_\star = 15.3\pm2.0$\,\gcc as reported by \citetalias{LibbyRoberts2023}).
This difference in impact parameter measurements likely also accounts for the difference in stellar inclinations, as the larger $b$ would also require a larger $i_\star$ to produce similar spot-crossing events given a polar spot.

\section{Conclusions}
\label{sec:conclusions}
In this paper, we used photometry from the \textit{Tierras} Observatory to show that TOI-3884 rotates with a period of \tierrasmcmcprot~days. We showed that the photometric variability of the star and the variations in transit shapes can be fit simultaneously by a starspot model with a large spot located near the pole and a planet on a misaligned orbit passing over this polar spot. Detailed analyses of spot-crossing events have previously been used to identify misaligned orbits for several planets \citep[e.g, HAT-P-11\,b, Kepler-63\,b, WASP-107\,b;][]{Sanchis-Ojeda2011,Sanchis-Ojeda2013,Dai2017}, but TOI-3884 is the first case where this technique has been applied to an M star host with a polar spot.

In principle, an observation of the spectroscopic transit of TOI-3884\,b to detect the Rossiter-McLaughlin effect would provide an independent measurement of the projected stellar obliquity, and break the degeneracy between the ``positive'' and ``negative'' pole solutions.
The projected stellar velocity implied by our best-fit stellar and planet orbital properties is $v\sin{i_\star} = 0.57\pm0.04$~\kms, which predicts a Rossiter-McLaughlin radial-velocity amplitude of approximately $8$~\ms for $\lambda_\star = 305^\circ$, detectable by current extreme-precision RV instruments.
However, the low impact parameter of the transit would severely limit the achievable precision on $\lambda_\star$ \citep{Albrecht2022}, so the photometric spot-crossing method will likely remain the most precise technique for characterizing the system geometry.

Our measurement of a high stellar obliquity for TOI-3884 confirms it as a member of the group of misaligned hot Neptunes around cool stars \citep[e.g.,][]{Bourrier2018,rmfit_Stefansson2022,Espinoza-Retamal2024}.
There are tentative hints that such planets may have a bimodal obliquity distribution, where the misaligned planets have a preference for nearly polar orbits \citep{Attia2023,Knudstrup2024}, as is the case for TOI-3884\,b.
Such misalignments could arise from primordial torquing of the protoplanetary disk due to magnetic interactions \citep{Lai2011}.
Alternatively, the planet's inclination may have been excited due to an exterior perturber causing Kozai-Lidov oscillations \citep[e.g.,][]{Fabrycky2007}, which would also cause inward migration onto a short-period orbit due to tidal friction.
\citet{Petrovich2020} more recently proposed that a secular resonance between an inner Neptune and an outer Jupiter could be encountered during disk dissipation and tilt the inner planet onto a polar orbit.
The latter two mechanisms require an exterior giant planetary or stellar mass perturber. Indeed, outer giant planets have been detected for a few polar Neptunes like HAT-P-11\,b \citep{Yee2018} and WASP-107\,b \citep{Piaulet2021}, with the right properties to trigger these dynamical processes \citep[e.g.,][]{Louden2024a,Yu2024,Lu2025a}.
In the case of TOI-3884\,b, no additional bodies are currently known in the system. Continued radial-velocity or astrometric monitoring may reveal such a perturber that could shed light on the formation history of this unusual planet.


The TOI-3884 system provides a benchmark for studying a polar starspot on an M dwarf. Assuming the spot is stable in time, future transit observations could build up a detailed map of the stellar pole as the transit chord crosses the spot at different angles. Additionally, changes in the transit shape over time will provide constraints on the latitude evolution of the spot. If the spot crossings dissipate over time, we will also be able to place a direct constraint on the spot lifetime.

\begin{acknowledgments}
The authors thank their anonymous referees for a helpful review that improved the quality of this work.

The \textit{Tierras} Observatory is supported by the National Science Foundation under Award No. AST-2308043. \textit{Tierras} is located within the Fred Lawrence Whipple Observatory; we thank all the staff there who help maintain this facility. S.W.Y. and J.G.-M. gratefully acknowledge support from the Heising-Simons Foundation. J.G-M. acknowledges support from the Pappalardo family through the MIT Pappalardo Fellowship in Physics.

The authors acknowledge Timothy Cunninghman for helpful conversations regarding the detection of the rotation period of TOI-3884 in archival ZTF photometry.
The authors also thank Rafa Luque for discussions about the RM effect for TOI-3884\,b, as well as Gudmundur Stefansson, Caleb Ca\~nas, Jessica Libby-Roberts, and Fei Dai regarding existing observations of TOI-3884.

This research made use of Lightkurve, a Python package for Kepler and TESS data analysis \citep{Lightkurve2018}.

Some of the data presented in this paper were obtained from the Mikulski Archive for Space Telescopes (MAST) at the Space Telescope Science Institute. The specific observations analyzed can be accessed via MAST \citep{MAST2021a, MAST2021b}.

This work makes use of observations from the LCOGT network. Part of the LCOGT telescope time was granted by NOIRLab through the Mid-Scale Innovations Program (MSIP). MSIP is funded by NSF.

\end{acknowledgments}





%
\facilities{ \textit{Tierras} Observatory, 
Transiting Exoplanet Survey Satellite (TESS),
Zwicky Transient Facility (ZTF),
Las Cumbres Observatory Global Telescope (LCOGT), FLWO:1.2m (KeplerCam)}

\software{ \texttt{starry} \citep{Luger2019};
          \texttt{scipy} \citep{scipy2020};
          \texttt{emcee} \citep{Fo13};
          \texttt{Lightkurve} \citep{Lightkurve2018};
          \texttt{AstroImageJ} \citep{Collins:2017}
          }



\bibliography{bib_file}{}
\bibliographystyle{aasjournalv7}



\end{document}

%% file: fit_results.tex
\startdata
&  \textbf{Sampled Physical Parameters}  & & \\ 
$P_\mathrm{rot}$ (days) & Stellar rotation period &  $\mathcal{N}(11.00,0.05)$ & 11.020$ \pm 0.015$\\ 
$i_\mathrm{\star}$ (deg.) & Stellar inclination &  $\mathcal{U}(0,90)$ & 22.3$^{+1.8}_{-1.6}$\\ 
$R_\mathrm{\star}$ ($R_\odot$) & Stellar radius & $\mathcal{N}(0.302, 0.012)$\tablenotemark{\scriptsize b} &0.306$ \pm 0.010$\\ 
$M_\mathrm{\star}$ ($M_\odot$) & Stellar mass & $\mathcal{N}(0.298, 0.018)$\tablenotemark{\scriptsize b} &0.295$^{+0.017}_{-0.018}$\\ 
$T_\mathrm{eff,\star}$ (K) & Stellar effective temperature & $\mathcal{N}(3180, 80)$\tablenotemark{\scriptsize b} &2985$\pm 65$\\ 
$T_\mathrm{eff,spot}$ (K) & Spot effective temperature & $\mathcal{N}(2900, 100)$\tablenotemark{\scriptsize b} &2791$^{+61}_{-72}$\\ 
$r_\mathrm{spot}$ (deg.) & Spot radius & $\mathcal{U}(22.5, 100)$ &31.2$^{+2.4}_{-1.9}$\\ 
$\lambda_\mathrm{spot}$ (deg.) & Spot latitude & $\mathcal{U}(-90, 90)$ & 80.5$\pm1.2$\\ 
$\phi_\mathrm{spot}$ (deg.) & Spot longitude at $t=0$ & $\mathcal{U}(0, 360)$ & 117.4$^{+4.9}_{-4.8}$\\ 
$s_\mathrm{spot}$ & Spot smoothing factor & $\mathcal{U}(0, \infty)$ &  0.169$^{+0.038}_{-0.043}$\\ 
$R_\mathrm{p}$/$R_\mathrm{\star}$ & Planet/star radius ratio & $\mathcal{N}(0.1970, 0.0020)$\tablenotemark{\scriptsize b} & 0.1906$ \pm 0.0022$\\ 
$i_\mathrm{p}$ (deg.) & Orbital inclination & $\mathcal{U}(0, 90)$; $\mathcal{N}(89.81, 0.18)$\tablenotemark{\scriptsize b} &89.934$^{+0.048}_{-0.094}$\\ 
$e$ & Orbital eccentricity & $\mathcal{N}(0.06, 0.06)$\tablenotemark{\scriptsize b} &0.042$^{+0.044}_{-0.029}$\\ 
$\Omega$ (deg.) & Longitude of periastron & $\mathcal{U}(0, 360)&180$^{+100}_{-130}$\\ 
$\lambda_\star$ (deg.) & Sky-projected stellar obliquity & $\mathcal{U}(0, 360)$ & 305.1$^{+5.4}_{-5.3}$\\ 
$u_{1, Tierras}$  & $Tierras$ first limb darkening coeff. & $\mathcal{N}(0.224, 0.200)$ & 0.409$^{+0.076}_{-0.070}$\\ 
$u_{2, Tierras}$  & $Tierras$ second limb darkening coeff.  & $\mathcal{N}(0.122, 0.200)$ & 0.281$ \pm 0.111$\\ 
$u_{1, g^\prime}$  & $g^\prime$ first limb darkening coeff. & $\mathcal{N}(0.4763, 0.2000)$ & 0.382$^{+0.197}_{-0.250}$\\ 
$u_{2, g^\prime}$  & $g^\prime$ second limb darkening coeff. & $\mathcal{N}(0.4000, 0.2000)$ & 0.752$^{+0.429}_{-0.384}$\\ 
$u_{1, i^\prime}$  & $i^\prime$ first limb darkening coeff. & $\mathcal{N}(0.2800, 0.2000)$ & 0.133$^{+0.167}_{-0.100}$\\ 
$u_{2, i^\prime}$  & $i^\prime$ second limb darkening coeff. & $\mathcal{N}(0.3293, 0.2000)$ & 0.489$^{+0.217}_{-0.229}$\\ 
\hline 
\hline 
&  \textbf{Sampled Nuisance Parameters}  & & \\ 
$b_{\mathrm{global},~Tierras}$ & Global $Tierras$ offset & $\mathcal{U}(-1, 1)$ &0.112$^{+0.014}_{-0.011}$\\ 
$m_{\mathrm{global},~Tierras}$ (days$^{-1}$) & Global $Tierras$ slope & $\mathcal{U}(-1, 1)$ &-6.3e-05$\pm+5.1e-06$\\ 
$b_{\mathrm{247},~Tierras}$ (ppt) & $Tierras$ offset night of Transit 247 & $\mathcal{U}(-1000, 1000)$ &0.72$^{+0.77}_{-0.79}$\\ 
$b_{\mathrm{254},~Tierras}$ (ppt) & $Tierras$ offset night of Transit 254 & $\mathcal{U}(-1000, 1000)$ &-0.20$^{+0.62}_{-0.67}$\\ 
$b_{\mathrm{258},~Tierras}$ (ppt) & $Tierras$ offset night of Transit 258 & $\mathcal{U}(-1000, 1000)$ &4.4$^{+0.6}_{-0.7}$\\ 
$b_{\mathrm{263},~Tierras}$ (ppt) & $Tierras$ offset night of Transit 263 & $\mathcal{U}(-1000, 1000)$ &7.5$ \pm 1.1$\\ 
$b_{\mathrm{272},~Tierras}$ (ppt) & $Tierras$ offset night of Transit 272 & $\mathcal{U}(-1000, 1000)$ &0.02$^{+0.91}_{-0.84}$\\ 
$b_{\mathrm{276},~Tierras}$ (ppt) & $Tierras$ offset night of Transit 276 & $\mathcal{U}(-1000, 1000)$ &8.5$^{+3.4}_{-3.3}$\\ 
$a_{\mathrm{276},~Tierras}$ & $Tierras$ airmass coeff. for Transit 276 & $\mathcal{U}(-1, 1)$ &-0.0107$\pm 0.0022$\\ 
$b_{\mathrm{flare},~Tierras}$ & $Tierras$ offset for flare on UT~2025~May~22 & $\mathcal{U}(-1, 1)$ & 38.0^{+3.6}_{-3.4} \\ 
$b_{\mathrm{258,~FLWO~1.2-m}}$ (ppt) & FLWO 1.2-m offset night of Transit 258 & $\mathcal{U}(-1000, 1000)$ &121$^{+26}_{-22}$\\ 
$b_{\mathrm{267,~FLWO~1.2-m}}$ (ppt) & FLWO 1.2-m offset night of Transit 267 & $\mathcal{U}(-1000, 1000)$ &190$^{+34}_{-31}$\\ 
$a_{\mathrm{258,~FLWO~1.2-m}}$ & FLWO 1.2-m airmass coeff. for Transit 258 & $\mathcal{U}(-1, 1)$ &0.0354$^{+0.0068}_{-0.0067}$\\ 
$a_{\mathrm{267,~FLWO~1.2-m}}$ & FLWO 1.2-m airmass coeff. for Transit 267 & $\mathcal{U}(-1, 1)$ &-0.006$^{+0.017}_{-0.018}$\\ 
$b_{\mathrm{267,~LCO~SSO~1-m}}$ (ppt) & LCO SSO 1-m offset night of Transit 267 & $\mathcal{U}(-1000, 1000)$ &123$^{+17}_{-14}$\\ 
$a_{\mathrm{267,~LCO~SSO~1-m}}$ & LCO SSO 1-m airmass coeff. for Transit 267 & $\mathcal{U}(-1, 1)$ &0.0001$^{+0.0015}_{-0.0014}$\\ 
\hline 
\hline 
&  \textbf{Derived Parameters}  & & \\ 
$c_{\mathrm{spot, \textit{Tierras}}}$ & Spot contrast in the \textit{Tierras} filter & - & 0.386\pm 0.038 \\ 
$c_{\mathrm{spot, \textit{g}^\prime}}$ & Spot contrast in the $g^\prime$ filter & -  & 0.608^{+0.047}_{-0.052} \\ 
$c_{\mathrm{spot, \textit{i}^\prime}}$ & Spot contrast in the $i^\prime$ filter & -  & 0.456^{+0.043}_{-0.044} \\ 
$c_{\mathrm{spot, TESS}}$ & Spot contrast in the TESS filter & -  & 0.383\pm 0.038 \\ 
$\Delta T$ (K) & Spot temperature difference & - & 198\pm 26\\
$a/R_\star$ & Scaled semi-major axis & - & 25.06^{+0.91}_{-0.71}\\
$b$ & Impact parameter & - & 0.029^{+0.041}_{-0.021}\\
$\rho_\star$ (g cm$^{-3}$) & Stellar density & - & 14.4^{+1.6}_{-1.2}\\
$\psi_\star$ (deg.) & True stellar obliquity & -  & 77.4^{+2.3}_{-2.5} \\ 
\enddata